\documentclass[Vancouver,Times2COL]{WileyNJDv5} 

\articletype{Research article}%

\received{Date Month Year}
\revised{Date Month Year}
\accepted{Date Month Year}
\journal{Journal}
\volume{00}
\copyyear{2023}
\startpage{1}

\usepackage{tabularx}
\usepackage{multirow}
\usepackage{cleveref}

\usepackage{tikz}
\usetikzlibrary{arrows.meta, shapes.geometric, arrows}
\tikzstyle{arrow} = [arrows = {-Stealth[scale=1.2]}]

\tikzstyle{text_box} = [rectangle, minimum width=1cm, minimum height=0.5cm, text centered, draw=black]
\tikzstyle{rounded_text_box} = [rectangle, minimum width=2cm, minimum height=0.5cm, rounded corners, text centered, draw=black]
\tikzstyle{large_rounded_text_box} = [rectangle, minimum width=2.5cm, minimum height=1.5cm, rounded corners, text centered, draw=black]
\tikzstyle{circ} = [circle, minimum size=0.3cm, text centered, draw=black]
\tikzstyle{bus} = [rectangle, minimum width=2cm,minimum height=0.15cm, draw=black, rounded corners]
\tikzstyle{block} = [rectangle, minimum width=1cm, minimum height=1.5cm, text centered, draw=black]
\tikzstyle{medium_rounded_text_box} = [rectangle, minimum width=1cm, minimum height=1cm, rounded corners, text centered, draw=black]

\usepackage{pgfplotstable}
\usepackage{pgfplots}
\pgfplotsset{compat=1.12}

\usepackage{acronym}
\acrodef{PINN}[PINN]{Physics-informed Neural Network}
\acrodef{PI-AC}[PI-AC]{Physics-informed Actor-Critic}
\acrodef{AC}[AC]{Actor-Critic}
\acrodef{ML}[ML]{Machine Learning}
\acrodef{PCC}[PCC]{Point of Common Coupling}
\acrodef{SMIB}[SMIB]{Single Machine Infinite Bus}
\acrodef{NN}[NN]{Neural Network}
\acrodef{VI}[VI]{Virtual Inertia}
\acrodef{GA}[GA]{Genetic Algorithm}
\acrodef{RL}[RL]{Reinforcement Learning}
\acrodef{DDPG}[DDPG]{Deep Deterministic Policy Gradient}
\acrodef{IBR}[IBR]{Distributed Energy Resources}
\acrodef{DSO}[DSO]{Distribution System Operator}
\acrodef{TSO}[TSO]{Transmission System Operator}
\acrodef{ROCOF}[ROCOF]{Rate of Change of Frequency}
\acrodef{PSO}[PSO]{Particle Swarm Optimizer}
\acrodef{IBR}[IBR]{Inverter-based Resource}
\acrodef{BESS}[BESS]{Battery Energy Storage System}

\begin{document}

\title{Physics-informed Actor-Critic for Coordination of Virtual Inertia from Power Distribution Systems}

\author[1]{Simon Stock}

\author[1]{Davood Babazadeh}

\author[1]{Sari Eid}
\author[1]{Christian Becker}

\authormark{Stock \textsc{et al.}}
\titlemark{Physics-informed Actor-Critic for Coordination of Virtual Inertia from Power Distribution Systems}

\address[1]{\orgdiv{Institute of Power and Energy Technology}, \orgname{Hamburg University of Technology}, \orgaddress{\state{Hamburg}, \country{Germany}}}

\corres{Corresponding author Simon Stock \email{simon.stock@tuhh.de} \\
\textbf{Author contributions}\\
Simon Stock: Conceptualization, Data curation, Investigation, Methodology, Resources, Software, Visualization, Writing – original draft, editing\\
Davood Babazadeh: Conceptualization, Investigation, Project administration, Supervision, Writing – review \& editing\\
Sari Eid: Data curation, Investigation, Software\\
Christian Becker: Conceptualization, Funding acquisition, Project administration, Resources, Supervision, Writing – review \& editing}

\fundingInfo{Publishing fees supported by Funding Programme Open Access Publishing of Hamburg University of Technology (TUHH).}

\abstract[Abstract]{The vanishing inertia of synchronous generators in transmission systems requires the utilization of renewables for inertial support. These are often connected to the distribution system and their support should be coordinated to avoid violation of grid limits. To this end, this paper presents the \acf{PI-AC} algorithm for coordination of \acf{VI} from renewable \acp{IBR} in power distribution systems.
Acquiring a model of the distribution grid can be difficult, since certain parts are often unknown or the parameters are highly uncertain. To favor model-free coordination, \ac{RL} methods can be employed, necessitating a substantial level of training beforehand.
The \ac{PI-AC} is a \ac{RL} algorithm that integrates the physical behavior of the power system into the \ac{AC} approach in order to achieve faster learning. To this end, we regularize the loss function with an aggregated power system dynamics model based on the swing equation. 
Throughout this paper, we explore the \ac{PI-AC} functionality in a case study with the CIGRE 14-bus and IEEE 37-bus power distribution system in various grid settings. The \ac{PI-AC} is able to achieve better rewards and faster learning than the exclusively data-driven \ac{AC} algorithm and the metaheuristic \ac{GA}.}



\maketitle

\renewcommand\thefootnote{}

\renewcommand\thefootnote{\fnsymbol{footnote}}
\setcounter{footnote}{1}

\section{Introduction}
\label{sec:introduction}
In recent years, renewable energy resources have replaced conventional generation to reduce the carbon footprint of energy production. Their dynamic characteristics are substantially different from conventional generation, since many of them are connected via power electronics, i.e. inverters. These lack inherent inertia in case of frequency events, in contrast to synchronous generators. This leads to higher vulnerability of the power systems, which can be partially compensated for by the immediate release of power from renewables via the inverter. The corresponding technique is termed \acf{VI}. 

Most renewable energy is connected to the distribution system, while conventional generation tends to be connected to the transmission system \cite{BDEW2017}. This can cause bidirectional power flows and violation of grid limits as a consequence. Thus, \acp{DSO} have been increasing the number of measurement devices in their grids to make full use of their operational capabilities and avoid structural grid extensions. These can be utilized in frameworks for the coordination of \ac{VI} in distribution grids \cite{DelNozal2020, stock_isie_23}. 

However, it is difficult to apply commonly used coordination algorithms, since there is often no distribution system model available. For that reason, model-based optimization can rarely be applied and the optimality of the corresponding results is questionable when inaccurate models are used. This puts focus on model-free techniques. There has been considerable work on metaheuristic approaches, such as the \acf{GA} and \acf{PSO} \cite{Magdy2019}. Although these approaches are able to optimize a given objective in a model-free manner, they do not obtain a strategy. Thus, they have to be rerun as soon as there are slight changes to the model or the objective. This brings up \ac{RL} techniques, such as \acf{AC}, which seek to obtain a coordination policy through offline learning. However, offline training still requires substantial amounts of time \cite{stock_isie_23}. 

In this paper, we strive to improve the training performance of an \acf{AC} approach that is model-free by design. To this end, we extend the \ac{AC} training loss by a physics-regularized term driven by a generic formulation for power system dynamics, the swing equation. This approach, the \ac{PI-AC}, is based on the idea that knowledge of the system can improve learning performance. Physics-based regularization has shown promising results in previous work applied to other \ac{ML} approaches \cite{Mahapatra2022, Karniadakis2021}. Overall, loss regularization is widely used in \ac{ML}, such as lasso and ridge regularization. 
We consider the \ac{PI-AC} to be model-free despite being driven by a physics-regularized loss, because the utilized physics formulation is generic for all power systems and does not require specific knowledge of the individual grid structure or setting parameters. 

In the literature, there exists considerable work on similar approaches, but most of them rely on an additional \ac{PINN} to operate as a transition model. 
In \cite{Mahapatra2022} a \ac{RL} approach is proposed that follows a similar idea. However, the swing equation is utilized in a \ac{PINN} transition model not as an additional regularization term. This approach increases the overall computational cost, since an additional \ac{PINN} model has to be trained. A similar approach is proposed in \cite{FARIA2024} for non-power system-related problems. 
In contrast to \cite{FARIA2024, Mahapatra2022}, the proposed \ac{PI-AC} seeks to regularize the loss function of the critic to influence the quality function. This does not cause considerable additional computation cost, since the number of trained networks remains the same.

In conclusion, this paper proposes the \ac{PI-AC} approach, which demonstrates superior performance with fewer training iterations compared to the \ac{AC} and \ac{GA} approach for coordination of \ac{VI} from distribution systems. 

This paper is structured as follows: \cref{sec:methodology} presents the problem formulation and in the following the \ac{PI-AC} methodology. It also introduces the fundamentals of the \ac{GA} which is used for comparison in this paper. This section is followed by the specification of the problem formulation for a power system \cref{sec:problem_statement}, which includes the relaxation of the given problem through penalty functions. In \cref{sec:case_study} we present the models that are used in the case study with the CIGRE 14-bus and IEEE 37-bus model and the corresponding equations. This section also presents the scenarios that will be used for evaluation in this paper. \Cref{sec:results} presents the results for all scenarios and compares the approaches. This section closes with a discussion on specific aspects that we found throughout the investigation. \Cref{sec:conclusion} closes this paper and concludes the main outcomes.  

\section{Methodology}
\label{sec:methodology}
In this paper, we strive to coordinate a dynamical system that can generally be described by a set of differential equations as follows 
\begin{align}
    \dot{\boldsymbol{x}}=g(\boldsymbol{u}, \boldsymbol{x};\boldsymbol{\lambda}).
    \label{eq:diff_eq}
\end{align}
The system states are described by $\boldsymbol{x}$ and its derivatives by $\dot{\boldsymbol{x}}$, $\boldsymbol{u}$ are the system inputs and $\boldsymbol{\lambda}$ the system parameters. Operator $g$ maps the inputs and parameters to the system states. To establish a model-free optimization problem, we describe the system as a black-box environment that produces a set of measurable states $\boldsymbol{s}\in\mathcal{S}$ under an applied set of actions $\boldsymbol{a}\in\mathcal{A}$. These states can be a subset of system states $\boldsymbol{x}$ and the actions $\boldsymbol{a}$ can be a subset of inputs $\boldsymbol{u}$. Actions are applied in order to optimize a given objective 
\begin{align}
    \min_{\boldsymbol{a}}  f(\cdot;\boldsymbol{s}).
    \label{eq:optimization_simple}
\end{align}
In this paper, we focus on the model-free coordination of \ac{VI} in power distribution systems, which can be set up in a form similar to \cref{eq:optimization_simple}. 
In the following, we will derive the \ac{PI-AC} which seeks to optimize the given objective $f$ through learning that is regularized by an approximation of \cref{eq:diff_eq} for faster learning. 
\subsection{Fundamentals of \acl{PI-AC}}
Most \ac{RL} techniques follow a model-free approach that considers the system as a black-box environment. To this end, we extend the previous formulation to a Markov decision process $(\mathcal{S}, \mathcal{A}, \mathcal{R}, p)$ with reward signal $r\in\mathcal{R}$ based on the objective $f$. The probability to transition from a given state and action to another state is given by $p$.
The goal of \ac{RL} is to find the optimal policy $\pi(\boldsymbol{s})$ to control the environment to obtain the best objective.\footnote{This is different for metaheuristic methods, which obtain a given set of parameters for the current environment and settings. They must retrain when significant changes are made in the environment or settings.} The policy learning itself is primarily guided by a quality function. This is learned separately using the reward $r$ based on the objective $f$. Thus, the policy learning is driven by the objective $f$.

This quality function can be represented by a \ac{NN} parameterized by a set of weights and biases $\boldsymbol{\Theta}^Q$. The \ac{NN} is called the critic, since it criticizes the actions taken by the current policy. The objective is incorporated into the learning of the quality function through the reward with $r=-f$
\begin{align}
    \mathcal{L}_{\text{critic}}=\mathbb{E}[(Q(\boldsymbol{s}_k,\boldsymbol{a}_k; \boldsymbol{\Theta}^Q)-(r+\gamma 
 Q(\boldsymbol{s}_{k+1},\boldsymbol{a}_{k+1}; \boldsymbol{\boldsymbol{\Theta}^Q})))^2].
    \label{eq:critic_loss}
\end{align}
Parameter $\gamma$ represents a discount factor that weights future expected rewards.
The parameters $\boldsymbol{\Theta}^Q$ are adapted based on $\mathcal{L}_{\text{critic}}$. The described approach is called the \ac{AC}. In the following, we extend it to the \ac{PI-AC}.

\textit{Physics regularized learning:}
In this paper, we aim to regularize the learning using the physical knowledge of the system. The goal is to achieve better performance, i.e. better objectives and faster learning. To this end, we augment the above equation \cref{eq:critic_loss} with a system model that approximates \cref{eq:diff_eq}
\begin{align}
    \mathcal{L}_{\text{physics}}=(\dot{\boldsymbol{x}}-d(\boldsymbol{u},\boldsymbol{x}; \hat{\boldsymbol{\lambda}}))^2.
    \label{eq:physics_loss}
\end{align}
This equation will approach zero if the estimates of the system parameters $\hat{\boldsymbol{\lambda}}$ are accurate. 
It is based on the measured system states $\boldsymbol{x}$ and a surrogate system model $d$ that utilizes the inputs $\boldsymbol{u}$ and the estimated system parameters $\hat{\boldsymbol{\lambda}}$. The estimates $\hat{\boldsymbol{\lambda}}$ are given by the critic. Note that $d$ approximates $g$, but does not necessarily has the same structure or rank. Instead, it can be a simplified representation of the system. 

Ultimately, this leads to the following critic loss formulation 
\begin{align}
    \mathcal{L}=\mathcal{L}_{\text{critic}}+\Phi \mathcal{L}_{\text{physics}}
    \label{eq:Loss_function}
\end{align}
with weighting factor $\Phi$.
The critic steers the learning of the policy, which itself can be represented by a \ac{NN} that is termed the actor. 

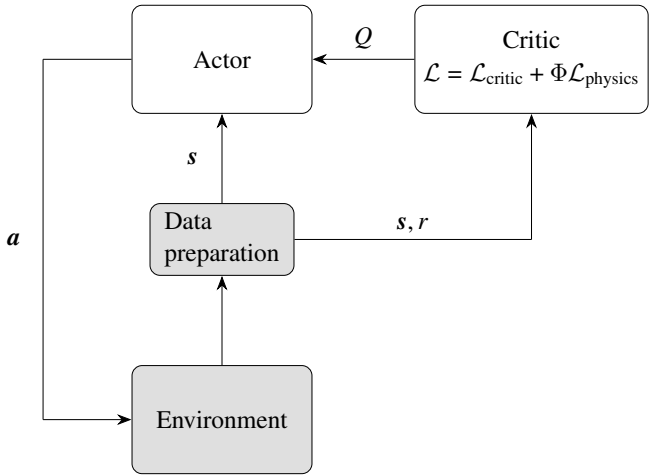
\begin{figure}
    \centering
    \resizebox{\linewidth}{!}{
    \begin{tikzpicture}[node distance=2.5cm]
        \node (Actor) [large_rounded_text_box]{Actor};
        \node (Critic) [large_rounded_text_box, right of=Actor, xshift=1.8cm, align=center]{Critic \\ $\mathcal{L}=\mathcal{L}_{\text{critic}}+\Phi\mathcal{L}_{\text{physics}}$};
        
        \node (PMU) [rounded_text_box, below of=Actor,  fill=gray!25, align=left]{Data\\ preparation};
        \node(Environment)[large_rounded_text_box, below of=PMU, fill=gray!25]{Environment};
        \coordinate[below of=Critic](misc1);
        \coordinate[left of=Actor, xshift=0.4](misc2);
        \coordinate[left of=Environment, xshift=0.4](misc3);

        \draw[arrow](Environment)--(PMU);
        \draw[arrow](PMU)--(Actor) node[midway, align=center, xshift=-0.4cm]{$\boldsymbol{s}$};
        \draw[](PMU)--(misc1) node[midway, align=center, yshift=0.2cm]{$\boldsymbol{s},r$};
        \draw[arrow](misc1)--(Critic);
        \draw[](Actor)--(misc2);
        \draw[](misc2)--(misc3) node[midway, align=center, xshift=-0.4cm]{$\boldsymbol{a}$};
        \draw[arrow](misc3)--(Environment);
        \draw[arrow](Critic)--(Actor) node[midway, align=center, yshift=0.3cm]{$Q$};
        
        %
    \end{tikzpicture}
    }
    \caption{Schematic representation of the proposed \ac{PI-AC} structure}
    \label{fig:PIAC_structure}
\end{figure}

We call the proposed approach the \acf{PI-AC} since it augments the \acf{AC} algorithm with a physics-informed loss term.\footnote{In contrast to the \ac{PINN} \cite{RAISSI2019}, we are not using the critic network to function as a surrogate model itself estimating the system states, but rather rely on the measurements of $\boldsymbol{x}$.} This structure is schematically shown in \cref{fig:PIAC_structure}.
The parameterized actor function $\mu(\boldsymbol{s}, \boldsymbol{\Theta}^{\mu})$ represents the current policy. This \ac{NN} is updated by applying the chain rule to the expected return from the initial distribution, with respect to the actor parameters. The resulting policy score function is calculated as follows  
\begin{align}
    \nabla_{\boldsymbol{\Theta}^{\mu}} J \approx \mathbb{E}[\nabla_{\boldsymbol{a}} Q(\boldsymbol{s},\boldsymbol{a}, \boldsymbol{\Theta}^Q)|\nabla_{\boldsymbol{\Theta}^{\mu}} \mu(\boldsymbol{s},\boldsymbol{\Theta}^{\mu})].
\end{align}
Note that this formulation relies on the quality function, i.e. the critic network, thus, policy learning is driven by the quality function and ultimately by the reward in \cref{eq:critic_loss}. 

Throughout this paper, we apply the \ac{DDPG} algorithm for learning. \acp{NN} are used to approximate both the actor and critic functions, which allows for learning in large state and action spaces. For efficient computation, learning can be performed in minibatches rather than online. This issue is addressed through a replay buffer. This buffer is a finite sized cache $\mathcal{R}$, which stores the tuple $(\boldsymbol{s}_k, \boldsymbol{a}_k, r_k, \boldsymbol{s}_{k+1})$ after environment exploration. Once the buffer has reached its maximum capacity, the oldest sample is discarded to make room for the new one. 
The \ac{DDPG} introduces duplicates of the actor and critic networks to ensure stable learning, namely $Q^{\prime}(\boldsymbol{s}, \boldsymbol{a}, \boldsymbol{\Theta}^{Q^{\prime}})$ and $\mu^{\prime}(\boldsymbol{s}, \boldsymbol{a}, \boldsymbol{\Theta}^{\mu^{\prime}})$, which are used to compute target values and are thus referred to as target networks. The network parameters $\boldsymbol{\Theta}^{Q^{\prime}}$ and $\boldsymbol{\Theta}^{\mu^{\prime}}$ are updated utilizing a soft update parameter $\tau$. This parameter slowly tracks them back to the learned networks $\boldsymbol{\Theta}^{\prime} \leftarrow \tau\boldsymbol{\Theta} +(1-\tau)\boldsymbol{\Theta}^{\prime}$. It is recommended to set $\tau << 1$ for stable convergence. The target values are limited to gradual modifications, which facilitates steady learning and, consequently, produces stable targets \cite{Lillicrap2015}. 
Exploring continuous action spaces can be difficult. Off-policy algorithms, such as \ac{DDPG}, are beneficial, as they decouple exploration and learning. The exploration policy can thus be build by adding random samples from a noise process following $\mathcal{N}$
\begin{align}
    \mu_{ex}(\boldsymbol{s}_k) = \mu(\boldsymbol{s}_k, \boldsymbol{\Theta}_k^{\mu}) + \mathcal{N}.    
\end{align}
For $\mathcal{N}$ an Ornstein-Uhlenbeck process is recommended for generation of temporally correlated exploration centered around 0.

\textit{Physics regularized learning for power system dynamics:}
In the context of power systems dynamics, the \ac{SMIB} model should be chosen as the surrogate model $d$ that regularizes learning in \cref{eq:physics_loss}. This model is a generic aggregated description of the dynamics of every power system, thus, the learning is still model-free, since no previous knowledge of the actual system is required. 
It can be described by \cite{Sauer2006}
\begin{align}
    \dot{\delta}&=\Delta\omega\\
    \Delta\dot{\omega}&=\frac{1}{2\hat{H}}(P_m-\hat{D}\Delta\omega-\hat{\frac{EV}{X_d}}\text{sin}(\delta-\varphi))
\end{align}
with system parameter estimates $\hat{\boldsymbol{\lambda}}=\{\hat{H}, \hat{D},\hat{\frac{EV}{X_d}}\}$. These represent the overall system inertia $H$, the damping $D$, the synchronous voltage $E$, the terminal voltage $V$ and the synchronous reactance $X_d$. The states of the system are the frequency deviation $\Delta \omega$ and the rotor angle $\delta$. The terminal voltage angle is described by $\varphi$. 
The parameter estimates $\hat{\boldsymbol{\lambda}}$ are determined by the critic, which therefore provides four outputs instead of one, namely the $Q$ and the system parameter estimates $\hat{\boldsymbol{\lambda}}$. This requires tuning the critic parameters $\boldsymbol{\Theta}^Q$, so the obtained $Q$ and $\hat{\boldsymbol{\lambda}}$ lead to a small overall loss \cref{eq:Loss_function}.

\subsection{Fundamentals of \acl{GA}}
The \ac{GA} is a metaheuristic optimization technique that seeks to minimize a given objective function. It is based on the fundamental concept that a group of system realizations, called the parents, are randomly placed throughout the search space. These are iteratively updated throughout the generations following genetic rules until convergence is reached. The set of parents can be expressed as
\begin{align}
    A = \{\boldsymbol{a}_1,\boldsymbol{a}_2, ...\boldsymbol{a}_N\}
\end{align} 
with $N$ parameters to influence the fitness, i.e. objective function $f(\cdot ; \boldsymbol{a}_i)$. After evaluating the corresponding fitness functions of the initial parameters, they are sorted after the fitness achieved. A high probability $p(\boldsymbol{a}_i)$ is assigned to those who achieve a high fitness. Mutation, replication, and crossover are employed to update the realizations and find the next generation \cite{brunton_kutz_2019}.

\section{Coordination framework and objective}
\label{sec:problem_statement}
\begin{figure}
    \centering
    \resizebox{\linewidth}{!}{
    \input{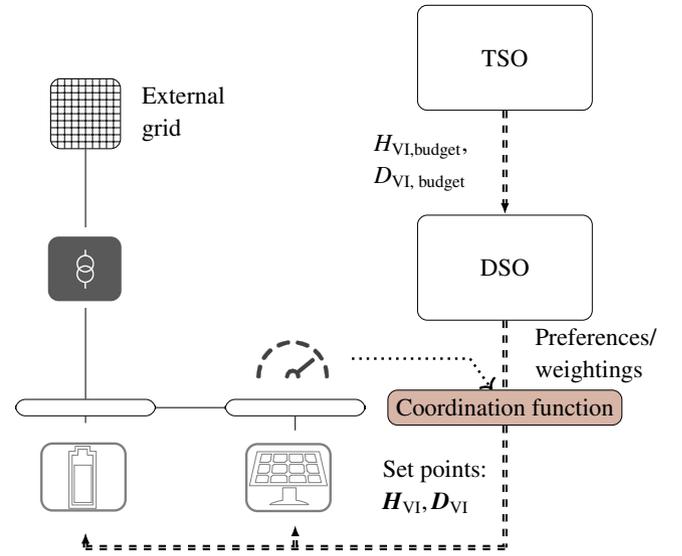}}
    \caption{The inertia support framework surrounding the coordination function}
    \label{fig:framework_schematic}
\end{figure}

In this paper, we strive to provide \ac{VI} from the distribution system for frequency support in a coordinated way. The superordinate framework was proposed in previous work \cite{stock_isie_23}. It is based on the assumption that the \ac{TSO} requests a certain amount of inertia that must be provided by the distribution system. The \ac{DSO} is obligated to coordinate the provision considering economic and system operational aspects. To this end, the \ac{VI} plant setpoints can be optimized using a certain objective. The framework can be concluded by the following steps: 
\begin{enumerate}
\item \ac{TSO} evaluates the system dynamics and determines the required amount of additional inertia and damping.
\item \ac{TSO} optimizes the inertia provision structure of the overall grid and assigns set points to specific regions.
\item \ac{DSO} receives the set points, i.e. inertia and damping budget for its region.
\item \ac{DSO} optimizes the provision considering certain preferences in its grid and assigns the corresponding set points to the \ac{VI} plants.
\end{enumerate}
This structure is also visualized in \cref{fig:framework_schematic}.

Often times, there is no accurate system model available of the distribution system. To this end, we apply the model-free \ac{PI-AC} approach. The optimization task can be written as follows, taking into account various \ac{VI} resources in the distribution system that can be controlled through their inertia and damping constants $H_{\text{VI},i}$ and $D_{\text{VI},i}$:
\begin{align}
\min_{H_{\text{VI},i}, D_{\text{VI}},i} f(C,\xi)
\label{eq:objective_wo_relaxation}
\end{align}
In this formulation, $C$ represents the economic costs and $\xi$ the system operational costs. The latter can be defined by the operator, for example, voltage deviation or transformer loading. 
In this paper, we set the economic costs to \cite{stock_isie_23} 
\begin{align}
    C=\sum c_i P_{\text{max},i}
\end{align}
and $P_{max,i}$ represents the maximum power provided by an individual plant considering the maximum expected \ac{ROCOF} and frequency deviation. These are weighted in the sum by an individual cost factor $c_i$, which is randomly chosen in the range of $0.5$ to $1.5$.
The system operational costs consist of the voltage deviation at and individual bus $V_i$ from the nominal value $V_{i,set}$, as follows
\begin{align}
    \xi=\sum c_{V,i}(|V_i|-|V_{i,set}|).    
\end{align}
We constrain \cref{eq:objective_wo_relaxation} by the given inertia and damping budget, the available maximum inertia and damping per plant and the voltage limits. This extends the previous formulation \cref{eq:objective_wo_relaxation}, so 
\begin{align}
\begin{split}
\min_{H_{\text{VI},i}, D_{\text{VI},i}} f(C,\xi)\\
\text{s.t.}\quad H_{\text{VI,sys}}\geq H_{\text{VI,budget}}\\
D_{\text{VI,sys}}\geq D_{\text{VI,budget}}\\
0\leq H_{\text{VI},i}< \frac{P_{\text{max},i}}{2\dot{\Delta \omega}_{\text{max}}}\\
0\leq D_{\text{VI},i}< \frac{P_{\text{max},i}}{\Delta\omega_{\text{max}}}\\
\Delta V_{i}\leq \Delta V_{\text{max}}
\end{split}
\end{align}
The \ac{VI} provided from the distribution system $H_{\text{VI,sys}}$ is calculated based on the sum of the individual contributions $H_{\text{VI},i}$ weighted by the corresponding nominal power and the system power. The damping is calculated in a similar way.
For this formulation, we assume that the maximum \ac{ROCOF} and maximum frequency deviation will not appear at the same time. 

The \ac{PI-AC} and \ac{RL} techniques in general are mostly guided by the reward function, thus, it is difficult to integrate constraints into the learning process. In this paper, we apply the penalty method, which relaxes the constraints by augmenting the objective with a corresponding penalty function. This additionally prevents the \ac{PI-AC} from being trapped in local minima that are close to the boundary of the feasible space. Allowing the algorithm to briefly step into the infeasible space tends to support broader exploration after all. Given that, the given constraints can be ensured in two ways. First, the penalty can be introduced earlier than the actual physical limit or constraint. Second, the constraints can be manually ensured, thus, a certain action is adapted after training to guarantee compliance with the given limits. In the following, we rely on the second approach, since the first does not guarantee compliance with the constraints. The third and fourth constraints are incorporated by limiting the action space in a corresponding way. All other constraints are used in the objective as penalties
\resizebox{\linewidth}{!}{
\begin{minipage}{\linewidth}
    \begin{align*}
    \begin{split}
    \min_{H_{\text{VI},i}, D_{\text{VI},i}} &f(C,\xi, p_{H_{\text{budget}}}, p_{D_{\text{budget}}}, p_{\Delta V_{\text{lim}}})\\
    \text{s.t.}&\\
    p_{H_{\text{VI,budget}}}&=\begin{cases} 0 &\text{if} \,(H_{\text{VI,budget}}-H_{\text{VI,sys}})\leq 0\\
    c_H \cdot (H_{\text{VI,budget}}-H_{\text{VI,sys}}) &\text{if}\, (H_{\text{VI,budget}}-H_{\text{VI,sys}})>0 \end{cases}\\
    p_{D_{\text{VI,budget}}}&=\begin{cases} 0 &\text{if} \,(D_{\text{VI,budget}}-D_{\text{VI,sys}})\leq 0\\
    c_{D} \cdot (D_{\text{VI,budget}}-D_{\text{VI,sys}}) &\text{if}\, (D_{\text{VI,budget}}-D_{\text{VI,sys}})>0 \end{cases}\\
    p_{\Delta V_i}&=\begin{cases} 0 &\text{if} \,(|\Delta V_i-\Delta V_{\text{max}}|)\leq 0\\
    c_{\Delta V} \cdot (|\Delta V_{i}-\Delta V_{i,\text{max}}|) &\text{if}\, (|\Delta V_i-\Delta V_{\text{max}}|)>0 \end{cases}
    \end{split}
    \end{align*}
\vspace{0.2cm}
\end{minipage}
}
The penalties are described by the corresponding penalty function $p$.
Ultimately, $f(\cdot)$ can be used as a reward in the reinforcement learning task with $r=-f(\cdot)$, so the reward can be maximized. 

\section{Case study}
\label{sec:case_study}
In this paper, we perform a case study based on the CIGRE 14-bus and IEEE 37-bus grid to explore the \ac{PI-AC} for coordination of multiple \ac{VI} sources. The superordinate transmission system is represented by an aggregated synchronous generator that is modeled with \cite{Sauer2006}
\begin{align}
    \dot{\delta}&=\Delta\omega\\
    \Delta\dot{\omega}&=\frac{1}{2H_{gen}}(P_m-D_{gen}\Delta\omega-\frac{EV}{X_d}\text{sin}(\delta-\varphi)\\
    \nonumber&+P_{gov})\\
    \dot{P}_{gov}&=-\frac{1}{T_s}(\Delta\omega+P_{gov}).
    \label{eq:generator_equations}
\end{align}
The inertia and damping of the generator are represented by $H_{\text{gen}}$ and $D_{\text{gen}}$. The synchronous longitudinal reactance $X_d$ and synchronous
voltage $E$ are coefficients to the electrical torque, while the terminal voltage $V$ is assumed to be 1. The corresponding terminal voltage angle is described by $\varphi$. Overall, the generator is driven by the mechanical power $P_m$. In addition, it provides a governor control through $P_{gov}$ with time constant $T_s$.

\begin{figure}
    \centering
    \resizebox{\linewidth}{!}{
    \begin{tikzpicture}[node distance=1.5cm]
        \node (Lf) [text_box,fill=black]{};
        \node (Rf) [text_box, left of=Lf, xshift=-0.2cm]{};
        \node (Lt) [text_box, right of=Lf, fill=black, xshift=0.2cm]{};
        \node (Rt) [text_box, right of=Lt]{};
        \node (meas) [rounded_text_box, above of=Lt,align=left, yshift=0.9cm, xshift=1.5cm]{Power\\meas.};
        \node (Pc) [rounded_text_box, left of=meas, xshift=-1.5cm]{P-control};
        
        \node (inv) [block, left of=Rf, align=left, xshift=-0.15cm]{Inverter};
        \node (emf) [rounded_text_box, above of=inv, align=left, yshift=0.9cm]{Virtual\\emf};

        \coordinate[above of=Pc, yshift=-0.75cm](misc12);
        \coordinate[below of=Pc, yshift=0.75cm](misc32);
        \coordinate[right of=Lf, xshift=-0.85cm](misc1);
        \coordinate[below of=misc1](misc2);
        \coordinate[left of=misc2, xshift=1cm](misc3);
        \coordinate[right of=misc2, xshift=-1cm](misc4);
        \coordinate[below of=misc2, yshift=1.2cm](misc5);
        \coordinate[left of=misc5, xshift=1cm](misc6);
        \coordinate[right of=misc5, xshift=-1cm](misc7);
        \coordinate[below of=misc5, yshift=1cm](misc8);
        \coordinate[left of=misc8, xshift=1.35cm](misc9);
        \coordinate[right of=misc8, xshift=-1.35cm](misc10);
        \coordinate[right of=Rt, xshift=-0.8cm](misc11);

        \coordinate[above of=misc1, yshift=-0.6cm](misc13);
        \coordinate[below of=meas](misc14);

        \node[above of=Rt, align=left, yshift=-1cm]{$R_T$};
        \node[above of=Lt, align=left, yshift=-1cm]{$L_T$};
        \node[above of=Rf, align=left, yshift=-1cm]{$R_f$};
        \node[above of=Lf, align=left, yshift=-1cm]{$L_f$};
        \node[above of=misc2, align=left, yshift=-1.2cm, xshift=-0.7cm]{$C_f$};

        \node(upe)[right of=inv, align=left, xshift=-0.7cm, yshift=-0.5cm]{};
        \node(loe)[right of=inv, align=left, xshift=-0.7cm, yshift=-2.3cm]{};
        \draw[->](upe)--(loe){};
        \node[right of=inv, align=left, xshift=-0.3cm, yshift=-1.5cm]{$\underline{E}$};

        \node(uVf)[right of=misc1, align=left, xshift=-0.8cm, yshift=-0.5cm]{};
        \node(lVf)[right of=misc1, align=left, xshift=-0.8cm, yshift=-2.3cm]{};
        \draw[->](uVf)--(lVf){};
        \node[right of=misc1, align=left, xshift=-0.3cm, yshift=-1.5cm]{$\underline{V}_f$};
        \node(uVt)[right of=Rt, align=left, xshift=-1cm, yshift=-0.5cm]{};
        \node(lVt)[right of=Rt, align=left, xshift=-1cm, yshift=-2.3cm]{};
        \draw[->](uVt)--(lVt){};
        \node[right of=Rt, align=left, xshift=-0.5cm, yshift=-1.5cm]{$\underline{V}_T$};

        \draw[dashed, line width=1pt, double distance=0.1pt](misc1)--(misc13);
        \draw[dashed, line width=1pt, double distance=0.1pt](misc13)--(misc14)node[midway, align=center, yshift=0.3cm]{$\underline{V}_f, \underline{I}_g$};
        \draw[dashed, line width=1pt, double distance=0.1pt,
             arrows = {-Latex[length=0pt 3 .5]}](misc14)--(meas);
        \draw[dashed, line width=1pt, double distance=0.1pt,
             arrows = {-Latex[length=0pt 3 .5]}](misc12)--(Pc)node[midway, align=center, xshift=0.4cm]{$\omega_{ref}$};
             \draw[dashed, line width=1pt, double distance=0.1pt,
             arrows = {-Latex[length=0pt 3 .5]}](misc32)--(Pc)node[midway, align=center, xshift=0.4cm]{$T_m$};
        \draw[](misc3)--(misc4);
        \draw[](misc1)--(misc2);
        \draw[](misc6)--(misc7);
        \draw[](misc5)--(misc8);
        \draw[](misc9)--(misc10);
        \draw[](inv)--(Rf);
        \draw[arrow](Rf)--(Lf)node[midway, align=center, yshift=0.25cm]{$\underline{I}_{RL}$};
        \draw[arrow](Lf)--(Lt)node[midway, align=center, xshift=0.1cm,yshift=0.25cm]{$\underline{I}_g$};
        \draw[](Lt)--(Rt);
        \draw[](Rt)--(misc11);
        \draw[ dashed, line width=1pt, double distance=0.1pt,
             arrows = {-Latex[length=0pt 3 .5]}](emf)--(inv) node[midway, align=center, xshift=0.5cm]{$e^*$};
        \draw[ dashed, line width=1pt, double distance=0.1pt,
             arrows = {-Latex[length=0pt 3 .5]}](Pc)--(emf);
        \draw[ dashed, line width=1pt, double distance=0.1pt,
             arrows = {-Latex[length=0pt 3 .5]}](meas)--(Pc)node[midway, align=center,yshift=0.4cm]{$-T_e$};


    \end{tikzpicture}}
    \caption{Simplified representation of the synchronverter \cite{Zhong2011}}
    \label{fig:Inv_diagramm}
\end{figure}
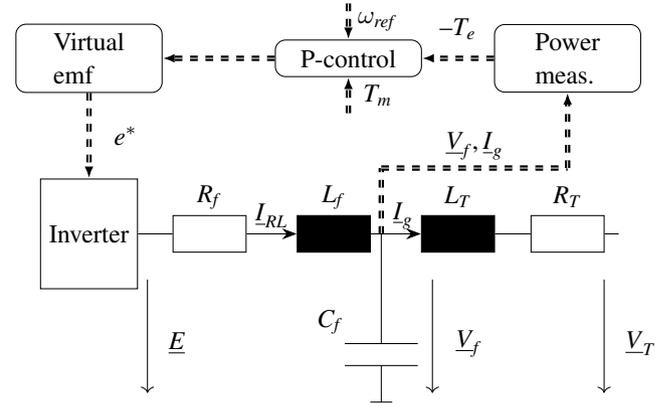

In this paper, we draw the energy for \ac{VI} provision from \acfp{BESS}, each \ac{BESS} has its own inverter, which is operated as a synchronverter \cite{Zhong2011}, as shown in \cref{fig:Inv_diagramm}.

\begin{figure}
    \centering
    \includegraphics[clip, width=\linewidth]{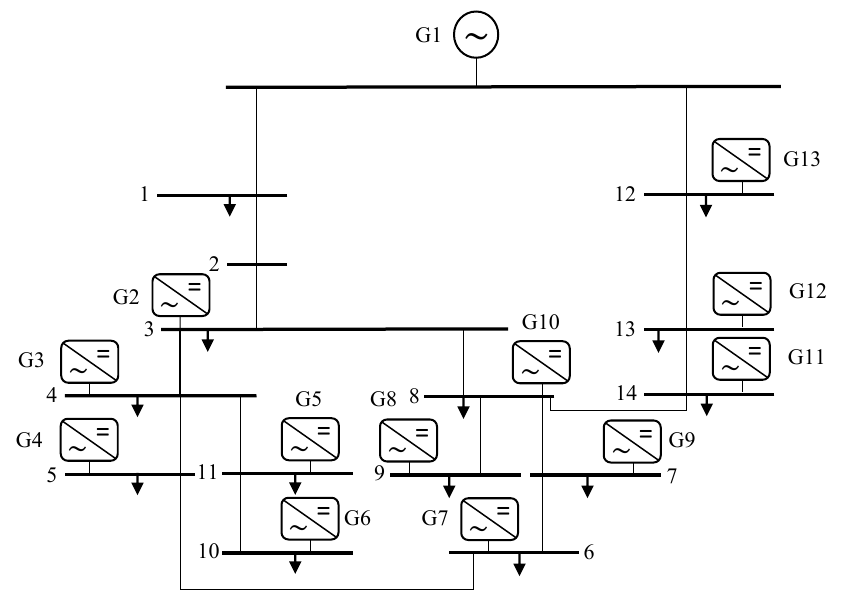}
    \caption{CIGRE 14-bus MV system structure}
    \label{fig:CIGRE_14_bus_system}
\end{figure}

\begin{figure}
    \centering
    \includegraphics[clip, width=\linewidth]{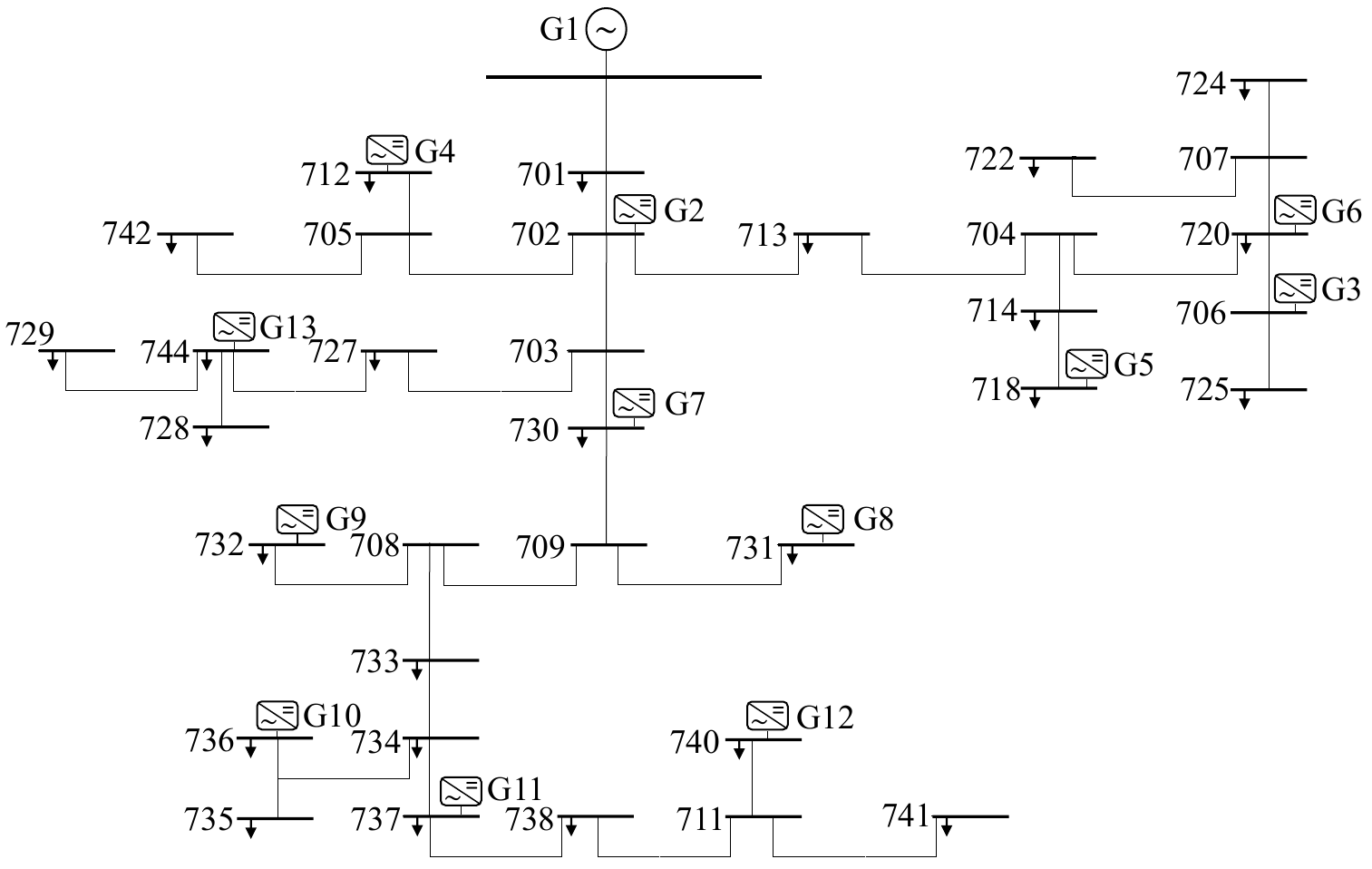}
    \caption{IEEE 37-bus MV system structure}
    \label{fig:IEEE37_bus_system}
\end{figure}

The RLC-filter and transformer ($R_T, L_T$) of the model follow the equations 
\begin{align}
    \dot{\underline{I}}_{RL}&=\frac{1}{L_f}(\underline{E}-\underline{V}_f-R_f \underline{I}_{RL}) \\
    \dot{\underline{V}}_f&=\frac{1}{C_f}(\underline{I}_{RL}-\underline{I}_g)\\
    \dot{\underline{I}}_g&=\frac{1}{L_t}(\underline{V}_f-\underline{V}_T-R_T \underline{I}_g).
\end{align}
The active power control follows the swing equation by calculating a virtual angular frequency $\omega_c$ and a virtual angle $\delta_c$ \cite{Zhong2011} 
\begin{align}
    \dot{\omega}_c&= \frac{1}{2H_c}(T_m-D_p (\omega_{\text{ref}}-\omega_c)-T_e)\\
    \dot{\delta}_c&= \omega_c\\
    e^*&=\dot{\delta}_c M_f i_f \text{sin}(\delta_c).
    \label{eq:synchronverter_equations}
\end{align}
The inertia and damping constants of the synchronverter are described by $H_c$ and $D_p$ respectively, the mechanical and electrical torque by $T_m$ and $T_e$, the mutual inductance by $M_f$ and the rotor excitation current $i_f$, which is assumed to be constant \cite{Zhong2011}. 
We place twelve \acp{VI}, i.e. \ac{BESS} with synchronverter, in both the CIGRE 14-bus and the IEEE 37-bus grid, as shown in \cref{fig:CIGRE_14_bus_system} and \cref{fig:IEEE37_bus_system}.

For investigation purposes, we simulate different grid situations, with varying shares of \ac{VI} and different load levels. The individual settings for each scenario are shown in \cref{tbl:scenarios_evaluation}, the transmission system inertia $H_{\text{TS}}$ is equal to the generator inertia in \cref{eq:generator_equations}. The budgets $H_{\text{VI,budget}}$ and $D_{\text{VI,budget}}$ are assigned to the individual \acp{VI} by the coordination function, which results in two vectors $\boldsymbol{H}_{\text{VI}}$, $\boldsymbol{D}_{\text{VI}}$ of 12 individual $H_{\text{VI},i}$, $D_{\text{VI},i}$. We measure the system states $\Delta \omega$ and $\delta$ at the generator and provide these as states for the \ac{PI-AC} training, so $\boldsymbol{s}_k=\{\Delta \omega, \delta\}$. The actions applied are the individual \ac{VI} parameters, so $\boldsymbol{a}_k=\{\boldsymbol{H}_{\text{VI}}, \boldsymbol{D}_{\text{VI}}\}$ with $H_{\text{VI},i}, D_{\text{VI},i}$ being the parameters for the $i$th plant. 
\begin{table}
\renewcommand{\arraystretch}{1.0}
\caption{Evaluation scenarios for test grids}
\label{tbl:scenarios_evaluation}
\centering
\resizebox{\linewidth}{!}{%
\begin{tabular}{l|rrrrr}
\toprule
Scenario &  $H_{\text{TS}}$ in p.u. & $D_{\text{TS}} $ in p.u. & $H_{\text{VI,budget}}$ in p.u. & $D_{\text{VI,budget}}$ in p.u. & $P_{\text{load}}$ in p.u.\\
\midrule
Reference & 1.1 & 0.8 & 13 & 13 &  1.0\\
\midrule
\multirow{2}{*}{$H_{\text{DS, budget}}$ variation} & 1.1 & 0.8 & 8 & 8 & 1.0  \\
&   1.1 & 0.8 & 18 & 18 & 1.0 \\
\midrule
\multirow{2}{*}{$H_{\text{TS}}$ variation} & 0.9 & 0.6 & 13 & 13 & 1.0 \\
& 1.8 & 1.5 & 13 & 13 &  1.0 \\
\midrule
\multirow{2}{*}{Load variation}
& 1.1 & 0.8 & 13 & 13 & 0.7\\
& 1.1 & 0.8 & 13 & 13  &  1.2\\
\bottomrule
\end{tabular}%
}
\end{table}
The power system simulations are run for $T=15\,\text{s}$ and step size $\Delta t=0.05\,\text{s}$ after reaching a stable operating point, so $\dot{\boldsymbol{x}}=0$. At that point a step change in the mechanical power of the generator is applied with $P_m=0.1\,\text{p.u.}$. This provokes a change in the frequency, thus the \acp{VI} that are placed in the distribution system react. 
The parameters of \ac{PI-AC} and \ac{DDPG} respectively are presented in \cref{tbl:DDPG_hyperparameter}. The Parameters for the benchmark systems, namely the 14-bus and 37-bus system, were taken from \cite{Strunz2014} and \cite{Rudion2006}.
\begin{table}[h!]
\renewcommand{\arraystretch}{1.0}
\caption{List of \acf{DDPG} Hyperparameters}
\label{tbl:DDPG_hyperparameter}
\centering
\resizebox{\linewidth}{!}{
\begin{tabular}{l|c}
\toprule
Hyperparameter & Value\\
\midrule
Optimizer & ADAM\\
Actor learning rate  & $1 \cdot 6^{-5}$\\
Critic learning rate  & $2 \cdot 6^{-5}$\\
Discount factor $\gamma$ & 0.995\\
Experience replay buffer $\mathcal{D}$ size & $300$\\
Target step size & 50\\
Repeat times & 10\\
Minibatch size & 50\\
Soft update coefficient $\tau$ & $2\cdot 10^{-4}$\\
Noise std $\sigma_{\text{Noise}}$ & $a_{max}$\\
Actor NN size (neurons) & $\text{dim}(\boldsymbol{s}_k)$; 100; $\text{dim}(\boldsymbol{a}_k)$\\
Critic NN size (neurons) & $\text{dim}(\boldsymbol{s}_k+\boldsymbol{a}_k)$; 100; 3\\
\bottomrule
\end{tabular}
}
\end{table}

\section{Results}
\label{sec:results}
We start this evaluation by comparing the overall reward achieved from the three approaches \ac{PI-AC}, \ac{AC} and \ac{GA}. \Cref{tbl:results_table} shows, that the \ac{PI-AC} achieves larger rewards in nearly all scenarios compared to \ac{AC} and \ac{GA}. This can be found for both grids, the 14-bus and 37-bus. It should be mentioned here, that the achieved rewards are not necessarily comparable between the systems, 14-bus and 37-bus, since the dynamics and therefore the achievable rewards are different.
In the results of the 14-bus system, we see that the \ac{PI-AC} always achieves the best rewards, except for when $H_{\text{DS, budget}}$ is high, meaning that high inertia and damping budgets are assigned. The largest difference can be found in the reference scenario and in high \ac{IBR} penetration, i.e. low $H_{\text{TS}}$. This is different for the 37-bus system, where the largest difference can be found in low $H_{\text{DS, budget}}$. Additionally, the results of the 37-bus system reveal that \ac{PI-AC} achieves the final reward faster than the \ac{AC} and \ac{GA} in nearly all cases. This is strongly pronounced in the 37-bus reference scenario and low $H_{\text{TS}}$. In the 14-bus system tests, it can also be seen that the \ac{PI-AC} achieves the final reward slightly faster in low $H_{\text{TS}}$ low compared to the reference scenario. This indicates that the higher share of \acp{IBR} in these scenarios, i.e. $H_{\text{TS}}$ low, results in faster learning of the \ac{PI-AC}. Overall, the difference in rewards between the \ac{PI-AC}, \ac{AC} and \ac{GA} is smaller in the 37-bus system. This is caused by the larger size of the 37-bus system and thus slower dynamics due to a smaller overall share of \acp{IBR}. 



\begin{table*}[h!]
\renewcommand{\arraystretch}{1.0}
\caption{Final reward $r_{final}$ and required iterations (it.) for coordination of CIGRE 14-bus grid and IEEE 37-bus grid}
\label{tbl:results_table}
\centering
\resizebox{\linewidth}{!}{%
\begin{tabular}{ll|rrrrrrrrrrrrrr}
\toprule
&& \multicolumn{2}{c}{Reference} &  \multicolumn{2}{c}{$H_{\text{DS, budget}}$ low} &  \multicolumn{2}{c}{$H_{\text{DS, budget}}$ high} &  \multicolumn{2}{c}{$H_{\text{TS}}$ low} &  \multicolumn{2}{c}{$H_{\text{TS}}$ high} &  \multicolumn{2}{c}{$P_{\text{load}}$ low} &  \multicolumn{2}{c}{$P_{\text{load}}$ high}  \\
Scenario & Algorithm & $  r_{\text{final}}$ &  it. &$  r_{\text{final}}$ &  it.&$  r_{\text{final}}$ &  it.&$  r_{\text{final}}$ &  it.&$  r_{\text{final}}$ & it.&$  r_{\text{final}}$ &  it.&$  r_{\text{final}}$ &  it. \\
\midrule
\multirow{3}{*}{14-bus} & PI-AC & -6.93 & 704 & -5.41 & 775 & -15.15 & 2277 & -6.39 &631 &-8.13 &676 &- 6.93 &704 &-6.93 &704 \\
& AC & -10.69 &662 &-6.48 & 991 &- 13.15 &1773 & -10.89 &775 &-10.45 &369 &-10.69 &662 &-10.69 &662 \\
& GA & -17.56 &910 &-11.21 &672 &-21.71 &398 &- 17.48 & 782 &-16.75 &910 & -17.09 &985 &-16.93 &833\\
\midrule
\multirow{3}{*}{37-bus} & PI-AC & -8.86 & 365 & -4.38 &352 &-12.51 &262&-7.62 &340 &-10.12&760 &-8.86 &365 &-8.86&365 \\
& AC & -9.42 &686 &-6.66 &583 &-12.49 &275 & -7.21 & 640 &-9.23 &459 &-9.42 &686 &-9.42 &686\\
& GA & -15.68 &928 & -14.51 & 981 & -20.99 & 919 & -15.01 &868 &-17.10 &760 &-16.45 &779 & -17.07 &527\\

\bottomrule
\end{tabular}
}
\end{table*}

\subsection{Comparison of \ac{AC} and \ac{PI-AC}}
\begin{figure}
    \centering
    \input{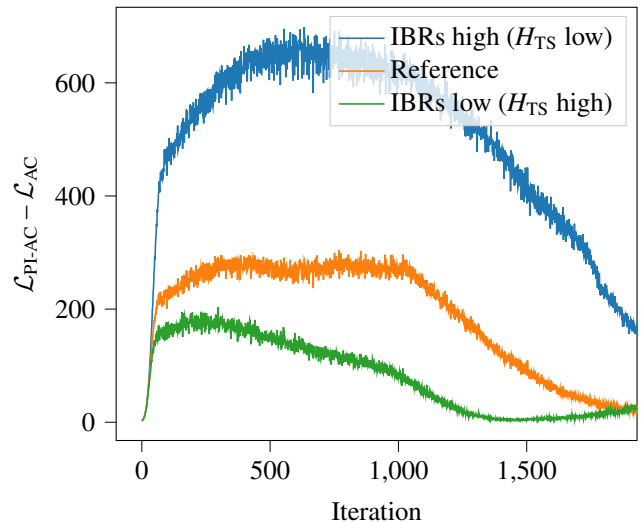}
    \caption{Critic loss difference of \ac{PI-AC} and \ac{AC} for different \ac{IBR} penetrations (14-bus system)}
    \label{fig:PIAC_AC_loss_comparison}
\end{figure}

\begin{figure}
    \centering
    \input{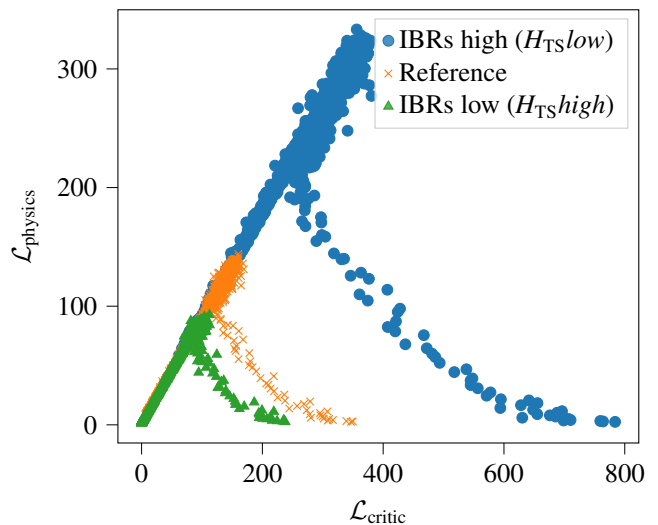}
    \caption{Influence of \ac{IBR} penetration on $\mathcal{L}_{\text{physics}}$ and $\mathcal{L}_{\text{critic}}$ (14-bus system)}
    \label{fig:physics_over_critic_loss}
\end{figure}

We continue this investigation by comparing the \ac{PI-AC} and the \ac{AC} in more detail. These are only different by the augmented loss function, as shown in \cref{eq:Loss_function}. \cref{fig:PIAC_AC_loss_comparison} reveals that the loss of \ac{PI-AC} is substantially greater. This stems from the additional physics regularization term. It can also be seen that the physics loss is much higher in scenarios with high shares of \acp{IBR}. This is caused by the deviation of the swing-equation-based regularization term from behavior of the real system. A large number of \acp{IBR} cause the system response to strongly deviate from the swing equation behavior, which is the basis for the physics-informed loss function in \ac{PI-AC}. This can also be seen in \cref{fig:physics_over_critic_loss}, which shows the physics regularization over the critic loss part. Furthermore, increasing \ac{IBR} penetration causes a combined growth of $\mathcal{L}_{\text{physics}}$ and $\mathcal{L}_{\text{critic}}$. 
This raises the general question of whether a more influential physics regularization in the loss function improves or hampers the learning of the algorithm.     
This question can be answered by comparing the reward of both \ac{PI-AC} and \ac{AC} for low and high $H_{\text{TS}}$ in \cref{tbl:results_table}. It can be seen that the reward in low $H_{\text{TS}}$, that is, a high share of \acp{IBR}, increases slightly compared to the reference scenario, while a higher $H_{\text{TS}}$, i.e., a low share of \acp{IBR}, leads to a significantly lower reward for \ac{PI-AC}. The rewards achieved by the \ac{AC} are comparable for both low and high \ac{IBR} penetration in the 14-bus system, thus, the differences can be assumed to result from the physics regularization term. However, the results from the 37-bus system show another influential factor. It is generally more difficult to obtain a better reward in the slower dynamics, i.e. $H_{TS}$ is high, since the influence of the actions, i.e. \ac{VI} parameters, cannot be clearly obtained. Thus, the \ac{AC} reward also decreases with slower dynamics, i.e. smaller share of \acp{IBR}, in the 37-bus case, this cannot be found for the 14-bus system, which has generally faster dynamics compared to the 37-bus system. 

\begin{figure*}
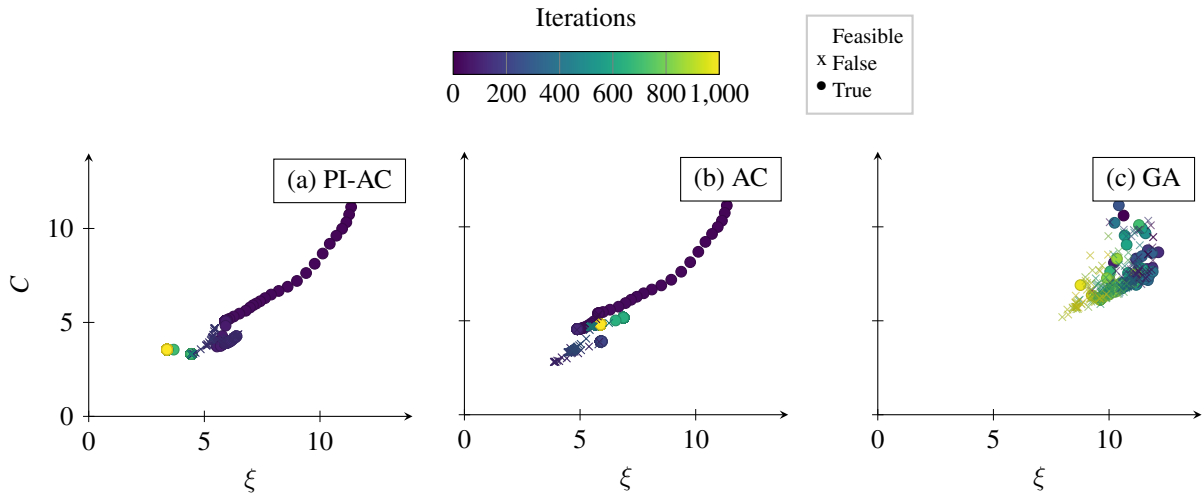

    \centering
    \begin{minipage}{\textwidth}
        \centering
        \begin{tikzpicture}

\begin{axis}[
    hide axis,
    scale only axis,
    height=0pt,
    width=0pt,
    colormap/viridis,
    colorbar horizontal,
    point meta min=0,
    point meta max=1000,
    colorbar style={
        title=Iterations,
        width=3.5cm,
        height=0.3cm,
    }]
    \addplot [draw=none] coordinates {(0,0)};
\end{axis}
\end{tikzpicture}
\hspace{0.45cm}
\begin{tikzpicture}

\definecolor{lightgray204}{RGB}{204,204,204}

\node[draw=lightgray204,thick]{%
\begin{tabular}{@{}r@{ }l@{}}
 \raisebox{2pt}{}&Feasible\\
 \raisebox{2pt}{x}&False\\
 \raisebox{2pt}{$\bullet$}&True \\
\end{tabular}};

\end{tikzpicture}
    \end{minipage}
    \raggedleft
    \begin{minipage}{0.3\textwidth}
        \vspace{0.5cm}
        \centering
        \input{img/PIAC_optimization_landscape_analysis_ref2.tikz}
        
    \end{minipage}
    \centering
    \begin{minipage}{0.3\textwidth}
        \centering
        \input{img/AC_optimization_landscape_analysis_ref.tikz}
        
    \end{minipage}
    \raggedright
    \begin{minipage}{0.3\textwidth}
        \centering
\begin{tikzpicture}

\definecolor{darkgray176}{RGB}{176,176,176}
\definecolor{lightgray204}{RGB}{204,204,204}
\definecolor{deepblue}{RGB}{21,76,121}
\pgfplotsset{colormap/viridis}
\begin{axis}[width=2.3in,
tick align=outside,
tick pos=left,
x grid style={darkgray176},
xlabel={\(\displaystyle \xi\)},
xmin=0, xmax=14,
xtick style={color=black},
y grid style={darkgray176},
ymin=0, ymax=14,
ytick style={color=black},
yticklabels={,,},
anchor=north,
yshift=-2cm,
axis lines = left,
]
\addlegendimage{empty legend}\addlegendentry{(c) GA}

    \addplot+ [scatter, only marks, mark=*, point meta=\thisrow{color}]
    table{meta=color%
    x  y color
10.6287	10.6012	0

10.2112	8.08891	28

11.64387	7.75458	42
12.1316	8.6463	43

11.87832	7.81179	50
11.17482	8.31239	51

11.27406	7.71645	54

11.67434	8.75508	56
11.04858	7.17371	57
11.64627	7.34815	58
11.04858	7.17371	59
11.04858	7.17371	60

11.211176	7.21258	73
11.74336	7.42185	74

11.41414	7.39285	76

11.38676	7.22636	78
11.37579	7.6546	79
11.7566	7.53598	80

11.33103	7.48009	82
11.17344	8.27248	83

11.04858	7.17371	88
11.54414	9.71721	89

10.4222	11.1456	92

11.15134	7.35393	95
11.10527	7.21	96

11.14065	7.25936	99

11.32071	7.37571	102
11.04858	7.17371	103
11.41776	8.43797	104
11.04858	7.17371	105
11.8543	8.60794	106
11.63329	7.30146	107
11.51963	7.68349	108
11.06065	7.1975	109

11.40797	7.5113	111
11.0165	7.1414	112
11.0165	7.1414	113

11.0165	7.1414	116

11.68019	7.1637	120

11.53924	7.22212	122
11.20629	7.41092	123
11.69641	7.35591	124

11.33846	7.83397	126

11.0165	7.1414	128

11.0165	7.1414	132
11.88753	7.61573	133

11.0165	7.1414	137
11.21349	7.37863	138
10.2379	10.21487	139

11.0165	7.1414	142
11.28528	7.1836	143

11.2028	6.95737	151
10.94215	7.06379	152

10.53839	6.76365	154
10.5424	6.75848	155
11.23576	7.53854	156
10.83548	6.90884	157
10.53839	6.76365	158
10.65826	7.00162	159
11.16977	6.91127	160

10.75213	7.03328	162
10.94204	7.06034	163
10.53839	6.76365	164
10.73392	6.86522	165

10.53839	6.76365	168

10.54943	6.71443	170
9.9948	7.31505	171

11.57787	9.65367	173
10.09022	6.66522	174

10.14439	6.68082	185

10.09062	6.66355	187
10.67264	9.53966	188
10.08916	6.65994	189

10.08916	6.65994	192
10.07518	6.65985	193
10.2582	6.87012	194

11.42449	9.91683	196
10.27236	6.75785	197
11.44816	7.76568	198
10.52686	7.00766	199

10.76072	9.051	203
10.07518	6.65985	204

10.82055	7.54578	208
10.09215	6.67259	209

10.09546	6.66853	216

10.07105	6.66538	220

10.28977	6.78178	225
10.49043	6.94147	226
10.07518	6.65985	227
10.07518	6.65985	228

10.04916	7.62542	230

11.0692	7.05256	236

10.4502	6.93498	238

11.29245	10.09605	243

10.07518	6.65985	247

10.07518	6.65985	249

9.64703	6.269	261

9.66821	6.27074	263
9.98877	6.52208	264

9.61972	6.12626	267

9.61972	6.12626	270

9.61972	6.12626	273

9.62265	6.12721	276
9.67143	6.17771	277

10.06094	6.47193	279

9.61972	6.12626	284

9.90809	7.26814	286
10.29931	6.64495	287

11.14666	7.48998	289
10.27887	6.54612	290

9.96959	6.37735	292

10.34083	8.30385	296

9.49544	6.21932	304

8.76373	6.8986	338

9.24974	6.35695	347

    };

    \addplot+ [scatter, only marks, mark=x, point meta=\thisrow{color}, fill opacity=0.3, draw opacity=0.6]
    table{meta=color%
    x  y color

11.80535	10.08079	1
11.90795	9.4536	2
11.90795	9.4536	3
11.69329	8.53134	4
11.68054	8.52779	5
11.83949	7.70821	6
11.89314	8.21163	7
11.27622	7.63291	8
11.26439	7.63962	9
11.72656	7.64118	10
11.18776	7.49215	11
11.27622	7.63291	12
11.87471	8.90853	13
10.95014	7.73786	14
11.58137	7.25976	15
10.30078	7.65738	16
11.18465	7.18946	17
10.6683	7.17424	18
11.39079	7.50771	19
10.0286	10.24555	20
11.39612	6.95192	21
11.64102	7.33672	22
11.26239	9.8369	23
11.30546	7.25504	24
11.58843	7.60619	25
10.96503	7.07498	26
11.12989	7.24572	27

11.33448	7.80832	29
10.94063	7.01437	30
11.26872	8.37547	31
9.26158	6.60684	32
11.12989	7.24572	33
11.035	7.9869	34
10.91561	7.47359	35
11.17155	7.45741	36
11.33626	9.86248	37
11.23666	7.58685	38
11.32998	7.74926	39
11.037	8.10428	40
11.19931	7.48077	41

11.44005	7.84584	44
11.04824	7.42321	45
11.09367	7.45967	46
11.05427	7.24004	47
11.17155	7.45741	48
11.15555	7.51508	49

11.0364	7.46302	52
11.5605	7.26882	53

10.72558	7.61966	55

11.69405	10.32498	61
10.03305	7.02459	62
11.0443	7.16845	63
11.04858	7.17371	64
10.44248	6.68081	65
10.81253	6.88654	66
10.80598	6.82475	67
11.14649	7.35382	68
10.35956	7.03628	69
11.15957	7.31395	70
10.5534	7.33248	71
11.03778	7.18561	72

10.95	6.88316	75

11.04455	7.16556	77

11.03075	7.06904	81

10.86746	6.95306	84
11.04589	7.18312	85
10.96878	9.49144	86
10.63527	9.84782	87

10.87427	7.4377	90
10.74991	7.10323	91

10.75011	6.83998	93
11.04049	7.06547	94

11.07531	9.82534	97
11.00647	7.0784	98

10.84608	7.12272	100
11.22505	8.15225	101

11.04622	7.17619	110

11.32702	7.17379	114
10.66084	6.7398	115

11.01021	7.09475	117
11.26904	7.12679	118
9.76274	6.36526	119

10.85701	7.10248	121

10.90234	7.27926	125

10.66695	6.79597	127

11.01269	7.00201	129
10.8938	7.11789	130
11.02231	7.15074	131

11.01598	7.14202	134
11.00457	7.94233	135
11.03688	7.14033	136

10.70075	6.84629	140
9.7916	6.81691	141

10.82516	6.80812	144
10.55881	7.05632	145
10.71788	6.80855	146
11.28046	8.15155	147
10.53517	6.75251	148
10.74318	9.49341	149
10.5584	6.76418	150

10.53517	6.75251	153

10.45322	6.52655	161

10.04646	6.7958	166
11.09416	9.68119	167

11.32853	9.02133	169

10.42447	6.45063	172

9.88202	6.60843	175
9.63069	6.14005	176
9.62638	6.49394	177
10.7571	7.4027	178
9.69809	6.44189	179
10.09022	6.66522	180
8.72719	6.12178	181
10.68087	7.91959	182
10.03137	6.39289	183
9.88083	6.78767	184

10.06616	6.57836	186

9.43079	6.87901	190
9.87256	6.37183	191

9.64221	6.27006	195

9.69055	6.54897	200
9.75755	6.57153	201
10.04522	6.61909	202

9.47178	6.22893	205
10.29129	6.78846	206
9.75074	6.76258	207

10.64312	9.70566	210
9.50035	6.40979	211
9.45392	6.29862	212
9.62699	5.89345	213
10.04781	6.78248	214
10.07518	6.65985	215

10.74386	6.7289	217
10.18019	8.53868	218
9.95496	6.39675	219

9.65631	6.85216	221
9.52411	6.96731	222
9.42382	7.25609	223
10.39814	6.77209	224

9.81747	6.3564	229

9.61769	6.12468	231
9.26597	6.38139	232
10.43556	7.99266	233
9.86363	7.30709	234
9.68423	6.53738	235

9.88549	6.20846	237

9.88499	6.28719	239
8.68991	6.08254	240
10.20237	6.48082	241
10.03286	6.83552	242

10.27311	6.89603	244
10.83384	8.62875	245
10.01478	6.53312	246

10.04248	6.51537	248

9.40153	6.24691	250
10.75319	6.98494	251
10.14222	6.7629	252
9.6374	6.53661	253
9.7989	6.32104	254
9.48875	6.9193	255
10.82242	8.22555	256
10.09752	6.66491	257
9.2896	6.20748	258
10.11446	6.98525	259
9.62336	6.32288	260

9.63274	6.22895	262

9.49654	6.11735	265
9.36403	6.2526	266

9.85166	6.79053	268
9.23508	6.0224	269

10.63483	6.53408	271
9.61791	6.13525	272

10.41626	8.69467	274
10.37064	6.42023	275

9.61939	6.17094	278

9.74757	6.99981	280
9.85703	6.19954	281
9.98019	6.32174	282
9.71554	6.19084	283

9.56429	6.42077	285

9.96999	9.50568	288

9.63008	6.22185	291

10.15786	7.28133	293
9.10333	5.73766	294
8.43499	5.66698	295

9.36564	6.06095	297
8.97463	7.31904	298
9.17601	6.60543	299
8.81041	5.88167	300
9.57321	7.70694	301
9.30028	6.41659	302
8.58366	5.73075	303

10.31419	7.17842	305
9.22099	6.19719	306
9.58185	6.17643	307
8.5899	5.61598	308
9.75556	6.20745	309
8.58538	5.7444	310
9.0441	6.92018	311
9.75656	6.22567	312
8.65003	5.62818	313
9.28008	7.68955	314
8.59123	5.95723	315
8.50767	5.72732	316
8.58366	5.73075	317
9.60333	6.17988	318
9.05192	7.01798	319
7.97714	5.19639	320
9.18781	6.62021	321
9.69424	7.93721	322
8.57678	5.92615	323
8.63917	5.99328	324
8.62032	5.73032	325
9.22512	7.00311	326
8.19765	5.5945	327
9.16048	6.18577	328
8.58394	5.73469	329
8.23365	5.33285	330
9.39512	5.99666	331
8.58396	5.7346	332
9.17987	6.12467	333
9.60371	6.18477	334
8.42266	5.70665	335
8.98634	7.12792	336
9.04721	6.6512	337

9.18459	8.47761	339
8.54051	5.74583	340
8.87775	6.00151	341
8.53819	5.96531	342
9.88099	6.41092	343
9.60292	7.64152	344
8.56044	6.3359	345
9.9573	8.51535	346

8.68439	5.69126	348
8.80678	5.91042	349

    };


\end{axis}

\end{tikzpicture}
        
    \end{minipage}
    \caption{Comparison of \ac{PI-AC}, \ac{AC} and \ac{GA} algorithms' movement through the optimization landscape (14-bus Reference scenario)}
    \label{fig:PIAC_GA_landscape_comparison}
\end{figure*}

\cref{fig:PIAC_GA_landscape_comparison} a, b compare the learning behavior of the \ac{PI-AC} and \ac{AC} in the $C$-$\xi$ plane. They indicate that the \ac{PI-AC} reaches the convergence region at an earlier point in time, i.e. at earlier iteration while the \ac{AC} moves through the plane at later iterations. A similar behavior has been found throughout this paper for most scenarios. It indicates that the physics regularization helps to drive the reward to the convergence region faster. 

\begin{figure*}
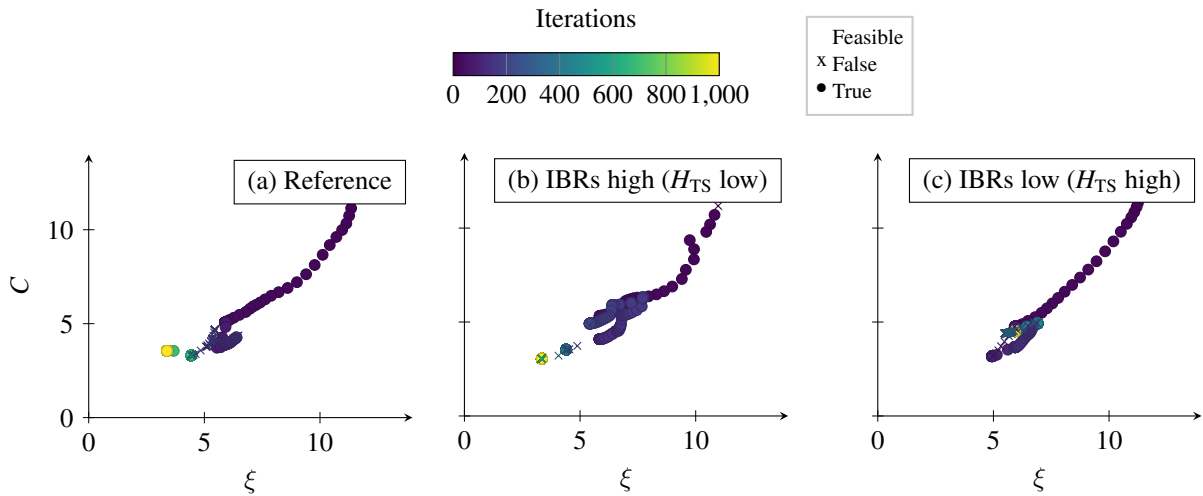

    \centering
    \begin{minipage}{\textwidth}
        \centering
        \begin{tikzpicture}

\begin{axis}[
    hide axis,
    scale only axis,
    height=0pt,
    width=0pt,
    colormap/viridis,
    colorbar horizontal,
    point meta min=0,
    point meta max=1000,
    colorbar style={
        title=Iterations,
        width=3.5cm,
        height=0.3cm,
    }]
    \addplot [draw=none] coordinates {(0,0)};
\end{axis}
\end{tikzpicture}
\hspace{0.45cm}
\begin{tikzpicture}

\definecolor{lightgray204}{RGB}{204,204,204}

\node[draw=lightgray204,thick]{%
\begin{tabular}{@{}r@{ }l@{}}
 \raisebox{2pt}{}&Feasible\\
 \raisebox{2pt}{x}&False\\
 \raisebox{2pt}{$\bullet$}&True \\
\end{tabular}};

\end{tikzpicture}
    \end{minipage}
    \raggedleft
    \begin{minipage}{0.3\textwidth}
        \centering
        \vspace{0.5cm}
        \input{img/PIAC_landscape_analysis/PIAC_optimization_landscape_analysis_ref.tikz}
        
    \end{minipage}
    \centering
    \begin{minipage}{0.3\textwidth}
        \centering
        \input{img/PIAC_landscape_analysis/PIAC_optimization_landscape_analysis_H_TS_low.tikz}
        
    \end{minipage}
    \raggedright
    \begin{minipage}{0.3\textwidth}
        \centering
        \input{img/PIAC_landscape_analysis/PIAC_optimization_landscape_analysis_H_TS_high.tikz}
        
    \end{minipage}
    \caption{Objective analysis considering the evolution in a $\xi$ and $C$ space over time for \ac{PI-AC} algorithm in a reference, high and low \ac{IBR} penetration scenario (14-bus system)}
    \label{fig:PIAC_optimization_landscape}
\end{figure*}

In the following, we investigate how different shares of \acp{IBR} and the corresponding losses influence the behavior of the \ac{PI-AC} in the optimization space, presented in the $C$-$\xi$ plane in \cref{fig:PIAC_optimization_landscape}. This figure shows that the increased \ac{IBR} penetration, low $H_{\text{TS}}$, leads to a stronger movement in the optimization landscape and ultimately results in lower costs. This is different for a lower share of \acp{IBR}, i.e. high $H_{\text{TS}}$. In this situation, the \ac{PI-AC} mostly moves around one spot and converges early with higher costs.

\subsection{Comparison of \ac{PI-AC} and \ac{GA}}
In this section, we evaluate the differences in the search strategies between the \ac{PI-AC} and the \ac{GA} algorithm. To this end, we compare the achieved costs for both algorithms in \cref{fig:PIAC_GA_landscape_comparison} a,c in the $C$-$\xi$ plane. It can be seen that the \ac{PI-AC} follows a straight path to the convergence region and spends the majority of iterations in that region to find the best setting. Focusing on the convergence region, it becomes visible that it is very shallow, due to the closely interwoven infeasible and feasible region. The \ac{GA} constantly moves in the direction of lower costs, however, the behavior is substantially different from the \ac{PI-AC}. The search is spread throughout the space, thus, the \ac{GA} settles in a much higher reward region than the \ac{PI-AC}.  
It should be highlighted that the \ac{PI-AC} learns a policy, while the \ac{GA} optimizes the problem at hand. To this end, the \ac{PI-AC} can quickly obtain results after training. In contrast, the \ac{GA} needs to be retrained for new results.

\subsection{Influence of the weighting factor $\Phi$}
The influence of the physics-based loss has been shown in previous results. We saw that higher physics loss often results in faster learning compared to the \ac{AC}. The higher physics loss was caused by different power system situations, where the increased share of \acp{IBR} led to an increased physics loss. 
These findings lead to a consecutive questions: does the influence of the physics loss saturate, so a higher loss reduces the achieved reward. This question will be evaluated in the following. 

\Cref{fig:Phi_heatmap} demonstrates that an increased weighting $\Phi$ strongly influences the reward and also changes the reward trajectory. It can be seen that the best rewards are achieved consecutively in the area around $\Phi=5000$. A higher loss leads to lower rewards. This is highly pronounced when $\Phi$ is larger than 100000. The rewards finally achieved with the different weightings are compared in \cref{fig:Phi_comparison}. This supports the results found in \cref{fig:Phi_heatmap}. Based on these findings, we set $\Phi=5000$ for all investigations.  

\begin{figure}
    \centering
    \includegraphics[clip, width=3.2in]{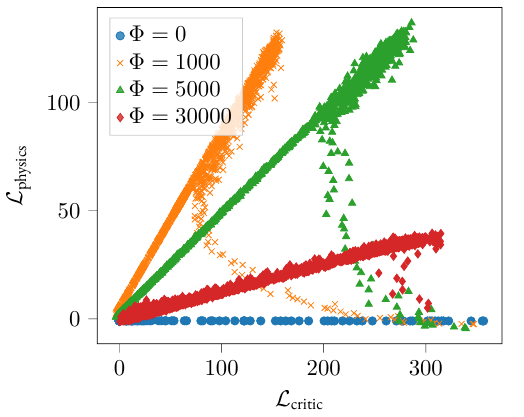}
    \caption{Comparison of different $\Phi$ weightings influence on $\mathcal{L}_{\text{physics}}$ and $\mathcal{L}_{\text{critic}}$(14-bus reference scenario)}
    \label{fig:PIAC_Phi_weightings_Q}
\end{figure}

In addition to its effect on the total loss $\mathcal{L}$, the physics loss also influences the critic loss $\mathcal{L}_{\text{critic}}$, shown in \cref{fig:physics_over_critic_loss}. This stems from the fact that the optimizer minimizes the total loss $\mathcal{L}$. Higher weighting shifts the focus to the physics loss, which allows a larger critic loss during training.  
In this case, a generally higher loss is not critical, since the parameter updates for $\boldsymbol{\Theta}$ of both \acp{NN} are performed based on the gradients $\nabla \mathcal{L}$. This has already been visualized in \cref{fig:physics_over_critic_loss}, but can also be seen in \cref{fig:PIAC_Phi_weightings_Q} for different $\Phi$. It can be seen here, that $\Phi=5000$ results in a loss trajectory that is close to the bisecting line, while a smaller weighting causes a higher physics loss and a much higher weighting a smaller physics loss. In the latter case, the optimizer partly neglects the critic loss to minimize the physics loss. 


\begin{figure}
    \centering
    \includegraphics[clip, width=3.2in]{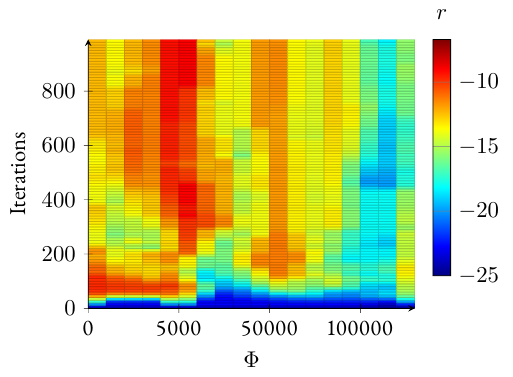}
    \caption{Reward trajectories during training for different $\Phi$ (14-bus reference scenario)}
    \label{fig:Phi_heatmap}
\end{figure}
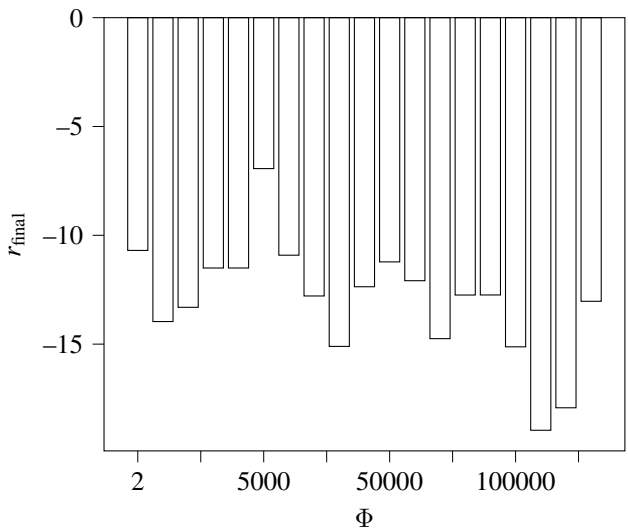
\begin{figure}
    \centering
\begin{tikzpicture}

\definecolor{darkgray176}{RGB}{176,176,176}
\definecolor{indigo68184}{RGB}{68,1,84}

\begin{axis}[
tick align=outside,
tick pos=left,
x grid style={darkgray176},
xlabel={\(\displaystyle \Phi\)},
xmin=-1.34, xmax=19.34,
xtick style={color=black},
xtick={-2.5,0,2.5,5,7.5,10,12.5,15,17.5,20},
xticklabels={,2,,5000,,50000,,100000,,},
y grid style={darkgray176},
ylabel={\(\displaystyle r_{\text{final}}\)},
ymin=-19.9057203569935, ymax=0,
ytick style={color=black}
]
\draw[] (axis cs:-0.4,0) rectangle (axis cs:0.4,-10.6943860615723);
\draw[] (axis cs:0.6,0) rectangle (axis cs:1.4,-13.9627986455185);
\draw[] (axis cs:1.6,0) rectangle (axis cs:2.4,-13.3085052263912);
\draw[] (axis cs:2.6,0) rectangle (axis cs:3.4,-11.5024791188042);
\draw[] (axis cs:3.6,0) rectangle (axis cs:4.4,-11.5021989775666);
\draw[] (axis cs:4.6,0) rectangle (axis cs:5.4,-6.934675883264);
\draw[] (axis cs:5.6,0) rectangle (axis cs:6.4,-10.91384519837);
\draw[] (axis cs:6.6,0) rectangle (axis cs:7.4,-12.7874116395393);
\draw[] (axis cs:7.6,0) rectangle (axis cs:8.4,-15.1048335287637);
\draw[] (axis cs:8.6,0) rectangle (axis cs:9.4,-12.3637259802376);
\draw[] (axis cs:9.6,0) rectangle (axis cs:10.4,-11.221855812807);
\draw[] (axis cs:10.6,0) rectangle (axis cs:11.4,-12.0862069487154);
\draw[] (axis cs:11.6,0) rectangle (axis cs:12.4,-14.7486944802776);
\draw[] (axis cs:12.6,0) rectangle (axis cs:13.4,-12.7442282607426);
\draw[] (axis cs:13.6,0) rectangle (axis cs:14.4,-12.7413326229933);
\draw[] (axis cs:14.6,0) rectangle (axis cs:15.4,-15.1235989698916);
\draw[] (axis cs:15.6,0) rectangle (axis cs:16.4,-18.9578289114224);
\draw[] (axis cs:16.6,0) rectangle (axis cs:17.4,-17.9249274739986);
\draw[] (axis cs:17.6,0) rectangle (axis cs:18.4,-13.0319282931634);
\end{axis}

\end{tikzpicture}
    \caption{Comparison of achieved rewards for different $\Phi$ (14-bus reference scenario)}
    \label{fig:Phi_comparison}
\end{figure}

\subsection{Discussion}
\label{subsec:discussion}
This chapter discusses additional thoughts on the proposed \ac{PI-AC} approach.

\subsubsection{Model-free operation}
In the beginning of this paper, we aimed to formulate a model-free coordination function. However, we utilize a generic system model, the swing equation, in the physics loss term. This can still be considered model-free since the swing equation formulation works equally as an aggregated model for all power systems, only the parameters change. These parameters are obtained by the critic network, thus, no prior knowledge about the system is required. 

\subsubsection{\ac{PI-AC} vs. \ac{GA}}
It has been found that the \ac{PI-AC} and the \ac{GA} apply different search strategies. The \ac{PI-AC} produces a learning trajectory that points in the direction of convergence and arrives in this area after a few iterations. On the contrary, \ac{GA} explores the space through its particles, causing a much more dispersed behavior. This ultimately causes a slower convergence.

\subsubsection{Runtime}
For runtime comparison, some calculations are run on an Intel i7 11700 CPU. The \ac{PI-AC} algorithm achieves an average runtime of $1425.86$ s for 100 training iterations. In comparison, the \ac{AC} algorithm yields similar runtimes, namely $1425.94$ s for 100 training iterations. It is evident that both algorithms have a comparable runtime. Hence, the extended loss function does not require  a noticeable extra computational effort. This stems from the fact that no additional \ac{NN} has to be trained.
The \ac{GA} achieves an average runtime of $4190$ s for 100 iterations. It has to rerun the simulation for every particle in each iteration, which makes training heavily dependent on the simulation time of the power system.
The GA does not acquire a policy that can be used after training for online execution. Instead, it must be rerun every time a parameter is altered. On the contrary, the PI-AC performs a complete training only once to discover the underlying policy of the problem. Subsequently, the policy can be applied as long as there are no substantial changes to the problem. When a retraining is necessary, the online optimization process can still be performed using the policy that was previously acquired.

\section{Conclusion}
\label{sec:conclusion}
In this paper, we presented the \acf{PI-AC} algorithm for model-free coordination of \acf{VI} provision from a power distribution system. The \ac{PI-AC} utilizes a physics regularization term in the \ac{RL} loss formulation driven by a generic representation of the power system dynamics, the swing equation. 
We show that the physics regularization improves the learning, the \ac{PI-AC} achieves better rewards in fewer iterations than the purely data-driven \acf{AC} in a case study based on the CIGRE 14-bus and IEEE 37-bus system. The results indicate that the physics-driven regularization is more pronounced in higher \ac{IBR} penetration and therefore leads to better results and superiority of the \ac{PI-AC} over the \ac{AC} in these scenarios. This is important to note for future inverter-dominated power systems. Throughout the case study, we also compare the \ac{PI-AC} to the metaheuristic \acf{GA} approach. The results show that the \ac{GA} takes significantly longer to converge and achieves worse rewards than the \ac{PI-AC}.

From the authors' perspective, the \ac{PI-AC} is not limited to the presented problem. Physics-based regularization in \ac{RL} can be a valuable tool for a variety of problems in power systems.

\bibliographystyle{WileyNJD-VANCOUVER}
\bibliography{library.bib}

\end{document}